\documentclass[conference]{IEEEtran}

\usepackage[dvipsnames]{xcolor}
\usepackage{amsmath,amsthm,amssymb,bbm}
\usepackage{algorithm}
\usepackage{algorithmicx}

\usepackage{array}
\usepackage{cite}
\newtheorem{theorem}{Theorem}
\newtheorem{lemma}{Lemma}
\newtheorem{proposition}{Proposition}

\usepackage{subfigure}
\usepackage{color}
\usepackage{graphicx}
\usepackage{flexisym}
\usepackage{dsfont}

\usepackage{algpseudocode}

\usepackage{mathtools}
\DeclarePairedDelimiter\ceil{\lceil}{\rceil}
\DeclarePairedDelimiter\floor{\lfloor}{\rfloor}
\ifCLASSINFOpdf
\else
\fi

\IEEEoverridecommandlockouts

\begin{document}
%
\title{Let's Share VMs: Optimal Placement and Pricing across Base Stations in MEC Systems}


%
%
%

\author{\IEEEauthorblockN{Marie~Siew$^{\dag}$, Kun~Guo$^{\dag}$, Desmond~Cai$^{\S}$, Lingxiang~Li$^{*}$, Tony~Q.S.~Quek$^{\dag}$}
\IEEEauthorblockA{$^{\dag}$Information Systems Technology and Design Pillar, Singapore University of Technology and Design\\
$^{\S}$Institute of High Performance Computing, Singapore\\
$*$University of Electronic Science and Technology of China, China\\
marie\_huilin@hotmail.com; guokun1218@foxmail.com; desmond-cai@ihpc.a-star.edu.sg;  
lingxiang.li@uestc.edu.cn;\\ tonyquek@sutd.edu.sg
}
\thanks{This work was supported in part by the National Natural Science Foundation of China under Grants 61901528, 62001254 and 61771263, and in part by the Hunan Natural Science Foundation under Grant 2020JJ5769. (Corresponding author: Kun Guo).}
}

\maketitle

\begin{abstract}
In mobile edge computing (MEC) systems, users offload computationally intensive tasks to edge servers at base stations. However, with unequal demand across the network, there might be excess demand at some locations and underutilized resources at other locations.
To address such load-unbalanced problem in MEC systems,  
in this paper we propose virtual machines (VMs) sharing across base stations. {\color{black}Specifically, we consider the joint VM placement and pricing problem across base stations to match demand and supply and maximize revenue at the network level.
To make this problem tractable, we decompose it into master and slave problems. For the placement master problem,} we propose a Markov approximation algorithm MAP
{\color{black}on the design of a continuous time Markov chain. As for the pricing slave problem, we propose OPA - an optimal VM pricing auction, where all users are truthful. Furthermore, given users' potential untruthful behaviors,}
we propose an incentive compatible auction iCAT along with a partitioning mechanism PUFF,
{\color{black}for which we prove incentive compatibility and revenue guarantees. Finally, we combine \textbf{MAP} and \textbf{OPA} or \textbf{PUFF} to solve the original problem, and analyze the optimality gap.} Simulation results show that collaborative base stations increases revenue by up to 50$\%$.

\end{abstract}


%
\IEEEpeerreviewmaketitle

\section{Introduction}
Mobile Edge Computing (MEC) is an enabler of exciting new technologies and applications like deep learning on devices, virtual and augmented reality, and smart city data analytics. These exciting new technologies and applications have high computation requirements. MEC enables them by allowing users to offload computationally intensive tasks to the network edge (e.g., base stations in cellular networks and access points in WiFi networks), which are equipped with computing capability by connecting to the edge servers \cite{mao2017survey}.
With servers placed at the network edge near the end users, Wide-area-network (WAN) delay is avoided, allowing it to meet the stringent latency requirements of delay sensitive tasks, that cloud computing is unable to \cite{hu2015mobile}.

Unlike cloud computing, the computational resources at the edge server are limited. Hence optimizing resource allocation in MEC is an important research question.
In particular, demand for computation is uneven across the network. Leading to excess demand at some coverage areas, and underutilized resources at others. {\color{black}In this load-unbalanced scenario,} there are users not being served, and from the network operator's perspective, resources are not efficiently utilized and revenue is not maximized. 
This prompts a global optimization and organization of resources over the network, to place resources more effectively in light of the network's demand pattern.

{\color{black}Virtual machine (VM) migration is perceived as a promising way to solve the load-unbalanced scenario \cite{mach2017mobile}.} There have been works on VM migration in MEC \cite{plachy2016dynamic,ksentini2014markov, ouyang2018follow, ma2020leveraging,wang2019delay}. These works investigate at the level of a single user, in response to user mobility. In contrast, there has been a lack of work from the global perspective.  
To this end, we propose the idea of “\textit{Collaborative Base Stations}”, where base stations share their VMs with each other. This involves the migration of VMs, in accordance with the relative demand across base stations. In particular, we consider a joint optimization of VM placement and pricing at base stations to match the demand and supply from the network level. {\color{black}A joint formulation is used because on one hand, the price at one base station has an impact on users' demand, which affects the VM placements. On the other hand, VM placement determines the resource supply at one base station. This way,} 
users' demand will be satisfied as much as possible and the revenue across the network is maximized.


However, some difficulties arise when solving the formulated joint VM Migration and Pricing for Profit maximization problem (\textbf{MPP}). Firstly, there is a sophisticated coupling of the price and VM placement variables, making it difficult to solve \textbf{MPP} directly. Secondly, \textbf{MPP} is a combinatorial optimization problem, with the number of VMs deployed at each base station being integers. It could be intractable, when the number of base stations increases and the total number of VMs deployed at the edge increases. {\color{black}Thirdly, the pricing at one base station is affected by the demand and bid information reported by the user. Users' potential untruthful behaviors make pricing at base stations challenging.} 

To tackle these difficulties, we first use primal decomposition to decouple the variables, decomposing \textbf{MPP} into {\color{black}the slave problem \textbf{NP} - Normalized Pricing problem, and master problem \textbf{VP} - VM Placement problem.} Next, we propose an online Markov approximation enabled algorithm which solves the \textcolor{black}{combinatorial \textbf{VP} in a distributed manner.}  
This helps to deal with the potential intractability when the problem size gets large.
It does so by modelling the different VM configurations as states of a Continuous Time Markov Chain (CTMC). 
The VM migrations happen according to the transition rate of the CTMC, which is in turn dependent on the performance level (revenue) of the placement configurations. How is the revenue of the VM placement configurations obtained? {\color{black}We solve \textbf{NP} to obtain the optimal revenue for each placement configuration. Specifically, at each base station we conduct either OPA - the Optimal Pricing Auction, or iCAT - an incentive CompAtible Truthful auction, which ensures users are truthful. iCAT guarantees the revenue $R$, when $R$ is less than or equal to the optimal.}
To successfully estimate $R$, we further present a user partitioning mechanism. The results of the auction will be fed back to \textcolor{black}{the base station and network operator,} 
directly influencing the transition rates of the CTMC.

Our contributions are summarized as follows:
\begin{itemize}
\item To deal with unequal demand across the MEC coverage areas, we formulate a joint VM migration and pricing problem {\color{black}across base stations to match demand and supply at the network level.} This works towards ensuring that user demand is met, resource placement is optimized globally, and the operator's revenue is maximized.
\item Due to 1) the combinatorial nature of the problem, 2) the coupling of price and placement variables, and 3) users having the incentive to hide their true valuations, we use primal decomposition to decompose the problem into a master and slave problem.
For the master VM placement problem, we present \textbf{MAP}, a Markov approximation-enabled algorithm which solves the combinatorial \textcolor{black}{problem in a distributed manner at individual base stations.}
\item To solve the pricing problem, we present an optimal pricing auction \textbf{OPA}, and prove that it is optimal. Besides, as users might have an incentive to hide their true valuations, we present an incentive compatible auction \textbf{iCAT}, prove that it is dominant strategy incentive compatible \textcolor{black}{and that its revenue is $R$, when $R$ is less than or equal to the optimal.} To estimate the target $R$, we present a user partitioning algorithm \textbf{PUFF}, \textcolor{black}{and prove that its competitive ratio is 4.}
\item We present the combined algorithm \textbf{cMAP} which solves our original joint VM placement and pricing problem, with an optimality gap of $\frac{1}{\beta}\log|\mathbb{V}|$. Following which, we conduct a perturbation analysis and show that the optimality gap of the stationary distribution caused by potential perturbations is bounded by $1-\exp(-2\beta \psi_{\text{max}})$, where $\psi_{\text{max}}$ is the perturbation error. 
\item Finally, we provide simulation results which show that our proposed solution \textbf{cMAP}: \textbf{MAP} + \textbf{OPA} converge to optimality, and analyze the impact of $\beta$. \textcolor{black}{While the performance of \textbf{cMAP}: \textbf{MAP} + \textbf{PUFF} is not optimal, it has a competitive ratio of $4$, as we have proved.} Results show that \textcolor{black}{our mechanism \textbf{cMAP} increases revenue by up to 50$\%$, compared to the baseline 
where base stations do not collaborate and VMs are not migrated.}
\end{itemize}

The rest of this paper is organized as follows. In Section \ref{sec:Related_Work}, we introduce related works. The system model and problem formulation are given in Section \ref{sec:System_Model}, which is followed by the optimal VM placement algorithm and the auction pricing algorithms in Sections \ref{sec:Placement_Algorithm} and \ref{sec:Pricing_Algorithm}. In Section \ref{sec:Algorithm_Analysis}, we give the complete implementation and analysis. In Section \ref{sec:Simulations} we discuss simulations results and in Section \ref{sec:Conclusions} we conclude.

\section{Related Works \label{sec:Related_Work}}
{\color{black}There are two mainstream ways to address the load-unbalanced problems for efficient resource utilization in MEC systems. On this basis we introduce the related works.

The first way is to optimize users' task offloading decisions, i.e. whether or not to offload, and which base station the user offloads to \cite{mao2017survey,mach2017mobile}. In this way, the computing resources at base stations are fixed and the users are handovered among base stations. For instance, \cite{Yuyi16, Xuchen16, dinh2017offloading} have optimized task offloading to strike a balance between energy consumption and delay from the perspective of users.
\cite{li2018energy} studied the static edge server placement problem.
\cite{QWangTSC19,YShihTSC19,kiani2017toward} aimed to maximize the network revenue through task offloading. 

This paper considers an alternative way, in which the computing resources are migrated among base stations to serve the associated users. Particularly, VM migration in MEC draws attention in industry and academic fields \cite{mach2017mobile,tao2019survey,wang2018survey}}. 
(Note that while there has been work on VM placement or migration for revenue maximization in clouds \cite{jiang2012joint,jennings2015resource}, these works are specific with respect to data center topologies.)
Most of the work on VM or service migration in MEC focus on improving user experience (e.g. reducing delay), in light of user mobility \cite{plachy2016dynamic,ksentini2014markov,ouyang2018follow,wang2019delay, ma2020leveraging}.
For example in \cite{plachy2016dynamic} Plachy \textit{et al.} proposed a dynamic VM placement and communication path selection algorithm.
In \cite{ksentini2014markov} Taleb \textit{et al.} optimized a policy on the service migration decision, given the user's distance. In \cite{ouyang2018follow}, Ouyang \textit{et al.} used Lyapunov optimization to optimize the placements over different timeslots.
Another line of research regarding VM migration looks at how it can maximize network profit or revenue. 
In \cite{sun2016primal}, Sun \textit{et al.} optimized the tradeoff between maximizing the migration gain and minimizing the migration cost.
In this work, we investigate from a novel perspective. We look at VM migration in MEC at a global level, in light of the network's demand patterns, for revenue maximization. And we formulate a joint VM migration and pricing problem because the price and migration decisions have a coupled impact on revenue. To the best of our knowledge, we do not know of many other works which take this approach.



Our proposed incentive compatible auctions and their proofs borrow from, but are different from the Profit Extractor and Random Sampling Auction in \cite{goldberg2001competitive,roughgarden2010algorithmic}. Profit Extractor and Random Sampling Auction cater to fully digital goods, with zero marginal cost of producing the next good, and hence an infinite supply. Unlike this, our network has a limited supply of VMs, resulting in unique novel algorithms and proofs.
\section{System Model and Problem Formulation\label{sec:System_Model}}
Consider an MEC system with {\color{black}$K$ base stations with heterogeneous computing capability}. Each base station $k$ is equipped with an edge server containing $v_{k}$ VMs. 
These are virtualised computing resources which users can offload their computationally intensive tasks to, at a price of $p_k$. 
Since the base stations are controlled by the same network operator, these VMs can be migrated from one base station to another, to optimize the utilization of resources. This global coordination of resources will help to deal with {\color{black}load-unbalanced scenarios} where there are excess demands in one coverage area, and underutilized resources in another part of the network. 

Each base station $k$ has a set of users $[1,...,i,..., n_k] \in U_k$ which are associated with it. Each user $i$ offloads its computationally intensive tasks to the edge server for auxiliary processing. Different users require different number of VMs, with user $i$ at base station $k$ requiring $r_{k,i}$ VMs. At base station $k$, different users respond differently to the price $p_k$.

A user $i$ at base station $k$ has willingness to pay $u_{k,i}$. The willingness to pay can be viewed as the utility a user gets from job computation using the VM. Different users have different willingness to pay. For example, a user with a more urgent job would have a higher willingness to pay than a user who is not as urgent. A user who will execute the job no matter what, with less regard of the price would have a higher willingness to pay (e.g., IoT sensors' periodic data analytics). A user will decide to execute its job if its payoff $\pi_{k,i}=u_{k,i}-p_k$ is {\color{black}non-negative}, i.e. if utility minus payment is greater than 0 $(\pi_{k,i}\geq 0)$. 
Therefore, the demand (total number of VM requests) at base station $k$ will be $\sum_{i\in U_k} r_{k,i} \mathds{1}_{\{u_{k,i}>p_k\}}$, where $\mathds{1}_{\{u_{k,i}>p_k\}}$ is the indicator function representing whether user $i$'s willingness to pay is higher than $p_k$.







The demand for VMs at each base station $k$ could be higher or lower than the supply $v_k$. 
Hence, the network operator would perform a global optimization of VMs, shifting them to locations with higher demand, to achieve a higher utilization of resources and to optimize its profit.
At the same time, the network operator sets prices $p_k$ differently across coverage areas, to obtain the highest possible revenue, in light of the varying demand across the network. 
{\color{black}The joint Migration and Pricing for Profit maximization problem (\textbf{MPP}) is as follows:
\begin{equation}
    \begin{aligned}
    \textbf{MPP}: \max_{\textbf{p},\textbf{v}} & \quad\sum_{k=1}^{K} p_k \min\left\{\sum_{i\in U_k} r_{k,i} \mathds{1}_{\{u_{k,i}\geq p_k\}}, v_k\right\}\\
     \text{s.t.} & \quad v_k \in \mathbb{Z}_0^{+}, k=1,...,K\\
    & \quad \sum_{k=1}^{K} v_k =V,
    \end{aligned}
\end{equation}
where $V$ is the total number of VMs, placed by the network operator across $K$ base stations. Besides, $\mathbb{Z}_0^{+}$ indicates the set of non-negative integers. In \textbf{MPP}, the decision variables are the prices across the various base stations $\textbf{p}=[p_1,..., p_k,...,p_K]$, in which each element is normalized (i.e., $p_k\in[0,1]$) without loss of generality,} and the VM placements across the network $\textbf{v}=[v_1,..., v_k,...,v_K]$. The objective function is the sum of the revenue obtained across base stations. In particular, it is the price multiplied by the number of units of demand which is met with supply. 

Some difficulties arise when solving \textbf{MPP}. Firstly, \textbf{MPP} is a combinatorial optimization problem, with $v_k$ being integers. It could be intractable, when the number of base stations increases and the total number of VMs increases. {\color{black}Even if we relax $v_k$ to continuous values, the problem is still non-convex.} Secondly, there is a coupling of $\textbf{p}$ and $\textbf{v}$ in the objective function, making it difficult to solve \textbf{MPP} directly. 

To tackle the difficulties in solving \textbf{MPP}, firstly we use primal decomposition \cite{Decomposition}, such that \textbf{MPP} is decomposed into {\color{black}slave problem \textbf{NP} - Normalized Pricing problem, and master problem \textbf{VP} - VM Placement problem. Specifically, fixing $\textbf{v}$, the slave problem is as follows:}
\begin{equation}
    \textbf{NP}:\max_{\textbf{p}} \quad \sum_{k=1}^{K} p_k \min\left\{\sum_{i\in U_k} r_{k,i} \mathds{1}_{\{u_{k,i}\geq p_k\}}, v_k\right\}. 
    \label{eq:Slave_Problem}
\end{equation}
Given the optimal solution from the slave problem, the master problem updates the VM migration decisions:
\begin{equation}
    \begin{aligned}
    \textbf{VP}: \max_{\textbf{v}} & \quad \Phi^*_{\textbf{v}}\\
    \text{s.t.} & \quad \textbf{v} \in \mathbb{V},
    \end{aligned}
\end{equation}
where $\Phi^*_{\textbf{v}}$ is the optimal value of $\textbf{NP}$ for the given $\textbf{v}$ {\color{black}and $\mathbb{V}=\{\textbf{v}|\sum_{k=1}^{K}v_k=V\bigcap v_k\in\mathbb{Z}_0^{+},k=1,...,K\}$ is the set of all possible VM placements across the network, with size $|\mathbb{V}|$.}

Following this, we propose a
\textcolor{black}{distributed Markov Approximation implementation to solve \textbf{VP}.} 
And finally, we propose both optimal and incentive compatible auction mechanisms to solve \textbf{NP}. 
We discuss the details in the following sections.

\section{The optimal VM placement algorithm\label{sec:Placement_Algorithm}}
\label{sec:MAP}
In this section, we will show how we solve the master problem \textbf{VP}. {\color{black}Particularly, we first reformulate and approximate \textbf{VP} and then,} propose a Markov approximation-enabled algorithm, named \textbf{MAP} - Markov Approx VM Placement algorithm. 

\subsection{Reformulating and Approximating \textbf{VP}}
The master problem \textbf{VP} can be rewritten as
\begin{equation}
    \begin{aligned}
    \textbf{VP-EQ}:\max_{\pi_{\textbf{v}}} &\quad \sum_{\textbf{v}\in\mathbb{V}} \pi_{\textbf{v}} \Phi^{*}_{\textbf{v}}\\
    \text{s.t.} 
    & \quad 0\leq \pi_{\textbf{v}}\leq 1,\forall \textbf{v}\in\mathbb{V}\\
    & \quad\sum_{\textbf{v}\in\mathbb{V}} \pi_{\textbf{v}} = 1,
    \end{aligned}
\end{equation}
where $\pi_{\textbf{v}}$ could be seen as the proportion of time spent in configuration $\textbf{v}$.

{\color{black}\textbf{VP} is an NP hard combinatorial optimization problem,} \textcolor{black}{and hence challenging to solve, even for a centralized implementation. Even if we relax $v_k$ to continuous values, the problem is still non-convex.}  
Therefore, we use the \textit{log-sum-exp} approximation $f(\Phi^*_{\textbf{v}})=\frac{1}{\beta}\log (\sum_{{\textbf{v}}\in\mathbb{V}} \exp(\beta \Phi^*_{\textbf{v}}))$ to approximate $\textbf{VP-EQ}$. \textcolor{black}{This approximation allows for a distributed implementation at individual base stations.
This is useful when the system dynamics change - when new users enter, or when users move from coverage area to area.} 
This approximation is upper bounded by $\frac{1}{\beta}\log|\mathbb{V}|$, following Proposition \ref{prop:optGap} \cite{boyd2004convex}:

\begin{proposition}
For $\beta>0$, we have 
\begin{equation}
    \max_{\emph{\textbf{v}}} \Phi^*_{\emph{\textbf{v}}} \leq \frac{1}{\beta} \log (\sum_{\emph{\textbf{v}}\in\mathbb{V}} \exp(\beta \Phi^*_{\emph{\textbf{v}}})) \leq \max_{\emph{\textbf{v}}} \Phi^*_{\emph{\textbf{v}}} + \frac{1}{\beta}\log|\mathbb{V}|.
\end{equation}
\label{prop:optGap}
\end{proposition}
Therefore, $\max_{\textbf{v}} \Phi^*_{\textbf{v}} = \lim_{\beta \rightarrow \infty} \frac{1}{\beta} \log (\sum_{\textbf{v}\in\mathbb{V}} \exp(\beta \Phi^*_{\textbf{v}}))$, i.e., the approximation tends towards $\textbf{VP-EQ}$ for large $\beta$. 
\textcolor{black}{As the \textit{log-sum-exp} function is a closed and convex function, the conjugate of its conjugate is itself, and hence we have $\frac{1}{\beta} \log (\sum_{\emph{\textbf{v}}\in\mathbb{V}} \exp(\beta \Phi^*_{\emph{\textbf{v}}}))=\sum_{\textbf{v}} \pi_{\textbf{v}} \Phi^{*}_{\textbf{v}} -\frac{1}{\beta} \sum_{\textbf{v}} \pi_{\textbf{v}} \log \pi_{\textbf{v}}$, according to \cite{boyd2004convex,chen2013markov}.}
Therefore the \textit{log-sum-exp} approximation of \textbf{VP-EQ} is equivalent to the following problem
\begin{equation}
    \begin{aligned}
    \textbf{VP-approx}:\max_{\pi_{\textbf{v}}}  & \quad \sum_{\textbf{v}} \pi_{\textbf{v}} \Phi^{*}_{\textbf{v}} -\frac{1}{\beta} \sum_{\textbf{v}} \pi_{\textbf{v}} \log \pi_{\textbf{v}}\\
    \text{s.t.} 
    & \quad 0\leq \pi_{\textbf{v}}\leq 1,\forall \textbf{v}\in\mathbb{V}\\
    & \quad\sum_{\textbf{v}\in\mathbb{V}} \pi_{\textbf{v}} = 1.
    \end{aligned}
\end{equation}
By solving the KKT conditions of \textbf{VP-approx}, the optimal solution is achieved in Theorem 1. 
\begin{theorem}
The optimal solution to \textbf{VP-approx} is 
\begin{equation}
    \pi_{\emph{\textbf{v}}}^*=\frac{\exp(\beta \Phi^*_{\emph{\textbf{v}}})}{\sum_{\emph{\textbf{v}} \in \mathbb{V}} \exp(\beta \Phi^*_{\emph{\textbf{v}}})}.
    \label{eq:OptDistrPiV}
\end{equation}
\end{theorem}

\textit{Proof.} Let $\lambda$ be the Lagrange multiplier associated with the constraint $\sum_{\textbf{v}\in\mathbb{V}} \pi_{\textbf{v}} = 1$. The Lagrangian of \textbf{VP-approx} will then be
\begin{equation}
    L(\pi_{\textbf{v}},\lambda) = \sum_{\textbf{v}\in\mathbb{V}} \pi_{\textbf{v}} \Phi^{*}_{\textbf{v}} -\frac{1}{\beta} \sum_{\textbf{v}\in\mathbb{V}} \pi_{\textbf{v}} \log \pi_{\textbf{v}} -\lambda(\sum_{\textbf{v}\in\mathbb{V}} \pi_{\textbf{v}}-1).
\end{equation}
Therefore, the KKT conditions will be: 
\begin{eqnarray}
    \Phi_{\textbf{v}}^* -\frac{1}{\beta}(\log  \pi_{\textbf{v}}^*+1)-\lambda = 0, \: \forall \textbf{v}\in\mathbb{V},\\ 
    \sum_{\textbf{v}\in\mathbb{V}} \pi_{\textbf{v}} = 1,\\ 
    \lambda\geq 0.
\end{eqnarray}
Solving the KKT conditions for the primal and dual optimal points $\pi_{\textbf{v}}^*$ and $\lambda^*$, we obtain $\pi_{\textbf{v}}^* = \exp(\beta( \Phi_{\textbf{v}}^*-\lambda)-1)$. Using the constraint $\sum_{\textbf{v}\in\mathbb{V}} \pi_{\textbf{v}} = 1$, we obtain $\lambda^* = \frac{1}{\beta} \log \sum_{\textbf{v}} \exp(\beta \Phi_{\textbf{v}}^* -1)$. Finally, substituting $\lambda^*$ into $\pi_{\textbf{v}}^* = \exp(\beta( \Phi_{\textbf{v}}^*-\lambda)-1)$, we obtain (\ref{eq:OptDistrPiV}). \qed

Therefore, by time-sharing among VM placement configurations according to the probability distribution $\pi_{\textbf{v}}^*$, we are able to solve \textbf{VP-approx}, and hence \textbf{VP-EQ}, \textbf{VP}, and \textbf{MPP} approximately.

\subsection{Solving VP: Algorithm design}
The idea consists of designing a Markov Chain, in which the state space is the space of possible VM placement configurations $|\mathbb{V}|$, and the stationary distribution is $\pi_{\textbf{v}}^*$, the optimal solution to \textbf{VP-approx}.
This would allow us to solve the joint VM placement and pricing problem \textbf{MPP }with an optimality gap of $\frac{1}{\beta}\log|\mathbb{V}|$.
To help us in the construction of the Markov chain, we use the following result from \cite{chen2013markov}:
\begin{lemma}
For any distribution of the form $\pi_{\textbf{v}}^*$ in (\ref{eq:OptDistrPiV}), there exists at least one continuous-time time-reversible ergodic Markov chain whose stationary distribution is $\pi_{\textbf{v}}^*$.
\end{lemma}

A continuous time-reversible markov chain (CTMC) is completely defined by its state space and transition rate. We let the state space be the space of possible VM placement configurations $\mathbb{V}$. 
The transition rate $q_{\textbf{vv\textprime}}$ indicates the rate at which the CTMC shifts from placement configuration \textbf{v} to \textbf{v\textprime}.
According to \cite{chen2013markov}, for the CTMC to converge to stationary distribution $\pi_{\textbf{v}}^*$, it needs to satisfy the following two conditions: 1) Irreducibility, meaning that any two states of the CTMC are reachable from each other. 2) Satisfaction of the detailed balanced equation: for any $\textbf{v},\textbf{v}\textprime  \in \mathbb{V}$, $\pi^*_{\textbf{v}} q_{\textbf{vv\textprime}}= \pi^*_{\textbf{v\textprime}} q_{\textbf{v\textprime v}}$. In other words, $\exp (\beta \Phi^*_{\textbf{v}}) q_{\textbf{vv\textprime}}= \exp (\beta \Phi^*_{\textbf{v\textprime}}) q_{\textbf{v\textprime v}}$ based on (\ref{eq:OptDistrPiV}).

Condition 1 can be satisfied because any two states (placement configurations) are reachable from each other. 
For Condition 2, let us set $q_{\textbf{vv\textprime}}=0$ for any two states which involve the migration of more than one VM from one base station to another. This is done to reduce the computation required, especially when the network is large.
For states which involve the migration of only one VM, we have
\begin{equation}
q_{\textbf{v}\textbf{v}'}= \exp( \frac{1}{2}\beta( \Phi^*_{\textbf{v}'}-\Phi^*_{\textbf{v}})).     
\label{eq:transitionEq}
\end{equation}
The detailed balance equation will be satisfied.
{\color{black}The transition rate $q_{\textbf{v}\textbf{v}'}$} is exponentially proportional to the performance of the target minus current VM placement configuration.
Therefore, when the performance (optimal revenue) of the target configuration is relatively higher than the current, there will be a higher transition rate, and vice versa.

The performance of each configuration $\textbf{v}$ is equivalent to its revenue obtained. In the next section, we show how to obtain the optimal revenue given a VM placement configuration $\textbf{v}$. In particular, we propose auction mechanisms to solve the slave problem \textbf{NP}. Following which, we will show how the algorithms solving the master problem \textbf{VP} and slave problem \textbf{NP} are combined to solve the original problem \textbf{MPP}.
\section{The Auction Pricing Mechanisms\label{sec:Pricing_Algorithm}}
In this section, we show how the slave problem \textbf{NP} can be solved. 
{\color{black}Specifically, \textbf{NP} defined in (\ref{eq:Slave_Problem}),} can be decomposed into individual pricing problems for each base station, where each base station $k$ solves the following problem:
\begin{equation}
    \textbf{NP-k}:\max_{p_k}  \quad p_k \min\left\{\sum_{i\in U_k} r_{k,i} \mathds{1}_{\{u_{k,i}\geq p_k\}}, v_k \right\}. 
    \label{revAtEachBS}
\end{equation}
{\textbf{NP-k}} can be solved by an auction. We provide two solutions, firstly \textbf{OPA} - Optimal Pricing Auction, which assumes the users are truthful, submitting bids $b_{k,i}$ equal to their true valuations $u_{k,i}$, and then \textbf{PUFF} - Partitioning Users For truthFulness {\color{black}mechanism, 
which includes an incentive CompAtible Truthful auction \textbf{iCAT}.} Our auction mechanisms are prior free, since they can be carried out without knowledge on the distribution of users' valuations $u_{k,i}$ . 

\subsection{The Optimal Pricing Auction (OPA)}
The mechanics behind \textbf{OPA} are as follows: {\color{black}users submit tuple $(r_{k,i}, b_{k,i})$ to base station $k$}, where $r_{k,i}$ is the amount of VMs requested by user $i$ at base station $k$, and $b_{k,i}$ is the bid indicating the user's willingness to pay for one VM. {\color{black}Since all users are truthful, the bid reported by the user is equal to its valuation (i.e., $b_{k,i}=u_{k,i}$). At price $p_k$, all users with valuation $u_{k,i} \geq p_k$ will be willing to participate in the auction. Then, we prove the optimal price will be $p_k^*\in \mathbb{B}_k=U_k$ in Theorem 2, where $\mathbb{B}_k$ and $U_k$ are the set of bids and valuations for users at base station $k$, respectively.} 




\begin{theorem}
\label{thr:optimalityOPA}
When all users are truthful, the optimal price of $\textbf{NP-k}$, termed as $p_k^*$, is found in $\mathbb{B}_k= U_k$.
\end{theorem}

\textit{Proof.} When all users are truthful, we have $\mathbb{B}_k=U_k$. Then, we prove that $p_k^{*}$ is found in $\mathbb{B}_k$. 

For the case with $p_k > \text{max}_{i\in U_k} b_{k,i} = \text{max}_{{i\in U_k}} u_{k,i}$, $\mathds{1}_{\{b_{k,i}\geq p_k\}} = \mathds{1}_{\{u_{k,i}\geq p_k\}} = 0$ holds, such that all users would reject to rent the VMs at base station $k$. Therefore, the revenue attained at base station $k$ is $\textit{Rev}(p_k) = p_k \min\{\sum_{i\in U_k} r_{k,i} \mathds{1}_{\{u_{k,i} \geq p_k\}}, v_k \} =0$.

Then, we analyse the case with $p_k < \text{max}_{i\in U_k} b_{k,i}$. Rearrange $\mathbb{B}_k$ in descending order {\color{black} and denote the set of ordered bids by $\{b_1,b_2,...,b_{n_k}\}$, where $b_i$ represents the $i$-th highest bid.} Using the fact that $b_{k,i}=u_{k,i}$, we have 
\begin{equation}
    \begin{aligned}
    \textit{Rev}(p_k=b_i-\epsilon) & = (b_i-\epsilon) \: \min\left\{\!\sum_{i\in U_k} r_{k,i} \mathds{1}_{\{b_{k,i}\geq (b_i-\epsilon)\}},v_k\! \right\}\\
    & < b_i \: \min\left\{\sum_{i\in U_k} r_{k,i} \mathds{1}_{\{b_{k,i}\geq b_i\}}, v_k \right\}\\
    & =  \textit{Rev}(p_k=b_i),
    \label{eq:Truth_Proof}
    \end{aligned}
\end{equation}
where $\epsilon<b_i-b_{i-1}$, no new users rent the VMs at base station $k$ by changing the price from $p_k=b_i$ to $p_k=b_i-\epsilon$, that is, $\min \left\{\sum_{i\in U_k} r_{k,i} \mathds{1}_{\{b_{k,i}
\geq (b_i-\epsilon)\}}, v_k \right\} =\min\left\{\sum_{i\in U_k} r_{k,i} \mathds{1}_{\{b_{k,i} \geq b_i\}}, v_k \right\}$ hold. Based on (\ref{eq:Truth_Proof}), we thus conclude that $p_k^{*}$ lies in $\mathbb{B}_k$. \qed

Using this insight that the optimal price belongs to the set of bids, the structure of our proposed \textbf{OPA} is summarized in Algorithm 1. In detail, after receiving the tuple $(r_{k,i}, b_{k,i})$ from all the users, base station $k$ will sort them into descending order with respect to $b_{k,i}$. For each unique bid $b_{k,i}$, the platform will set $\bar{p}_k=b_{k,i}$, and calculate the revenue $\textit{Rev}(\bar{p}_k) = \bar{p}_k \min\{\sum_{i\in U_k} r_{k,i} \mathds{1}_{\{u_{k,i}\geq \bar{p}_k\}}, v_k \} $. Following which, it will optimize over $\bar{p}_k$ and achieve $p_k^*= \text{argmax}_{\bar{p}_k=b_{k,i},\forall i\in U_k} \textit{Rev}(\bar{p}_k)$. 

\begin{algorithm}
\caption{OPA: Optimal Pricing Auction}\label{algo:OPAauction}
\begin{algorithmic}[1]
\State \textbf{Input:} Tuple $(r_{k,i}, b_{k,i}),\forall i\in U_k$
\State Sort $(r_{k,i}, b_{k,i})$ according to descending order with respect to $b_{k,i}$. 

\For{all unique $b_{k,i}$}
\State Set $\bar{p}_k= b_{k,i}$
\State $\textit{Rev}(\bar{p}_k) \gets \bar{p}_k \min\{\sum_{i\in U_k} r_{k,i} \mathds{1}_{\{u_{k,i} \geq \bar{p}_k\}}, v_k \} $ \Comment{By Eq. (\ref{revAtEachBS})}
\EndFor

\State \textbf{Output:} $p_k^* \gets \text{argmax}_{\bar{p}_k=b_{k,i},\forall i\in U_k} \textit{Rev}(\bar{p}_k)$

\State \textbf{end}
\end{algorithmic}
\end{algorithm}
\subsection{The Incentive CompAtible Truthful Auction (iCAT)}
In reality, users may have an incentive to submit bids unequal to their true valuations (i.e. $b_{k,i} \neq u_{k,i}$), hoping to achieve a higher payoff. {\color{black}Therefore, we present incentive compatible auction mechanism \textbf{iCAT}, 
by which the user's dominant strategy is to be truthful.}

Given a target revenue $R$, the auction mechanism will post price $p_k=\frac{R}{\min\{\sum_{i \in U_k} r_{k,i}, v_k \}}$, where $\sum_{i \in U_k} r_{k,i}$ is the total demand of the users currently in the auction. Users will decide whether or not to accept the offer by weighing if their payoff $p_k-u_{k,i}$ {\color{black}is not lesser than} $0$ (individual rationality met). If any user $i$ rejects the offer, he is removed from future rounds of the auction. {\color{black}Then, the set of users in the auction is updated as $ U_k \gets U_k \setminus \{ i\}$. The process repeats: base station $k$} obtains the new demand $\sum_{i \in U_k} r_{k,i}$ of users currently in the auction, and broadcasts the new price $p_k=\frac{R}{\min\{\sum_{i \in U_k} r_{k,i}, v_k \}}$.
If all users remaining in the auction 
accept the offer, they will be the winners, paying the last offer price $p_k$.
Therefore, base station $k$ would rent $\min\{\sum_{i} r_{k,i} \mathds{1}_{\{u_{k,i}\geq p_k\}}, v_k \}$ units of VMs to users with bids in the set $U_k $ at price $p_k=\frac{R}{\min\{\sum_{i\in U_k} r_{k,i} \mathds{1}_{\{u_{k,i}\geq p_k\}}, v_k \}}$. 

{\color{black}The complete iCAT is summarized in Algorithm 2.} The main idea behind this mechanism is that it prunes the set of auction users until it obtains a set $U_k$ where: the users in $U_k$ are willing to pay $p_k=\frac{R}{\min\{\sum_{i\in U_k} r_{k,i} ,v_k\}}$, the price at which the base station obtains revenue $R$ given demand $\sum_{i\in U_k} r_{k,i}$.
Note that our auction mechanism does not involve the users submitting any bids $b_{k,i}$. Truthfulness is ensured via the structure of the mechanism, as proved in Theorem \ref{ICMechanism}. \textcolor{black}{In particular, we prove that \textbf{iCAT} is dominant strategy incentive compatible, meaning that being truthful gives the users a higher payoff compared to any other strategy.} 

\begin{algorithm}
\caption{iCAT: incentive CompAtible Truthful Auction}\label{algo:TruthfulAlgo}
\begin{algorithmic}[1]
\State \textbf{Input:} {\color{black}Initialize $U_k$, the number of VMs required by user $i$ ($r_{k,i}$), and target revenue $R$ at base station $k$.}
\While{{\color{black}$U_k$ is not empty}}
{\color{black}\State Base station $k$ posts price $p_k=\frac{R}{\min\{\sum_{i \in U_k} r_{k,i}, v_k \}}$;

\If{$u_{k,i}<p_k$ for any user $i\in U_k$}
    \State User $i$ rejects to join in the auction;
    \State Base station $k$ updates $U_k \gets U_k \setminus\{ i \}$;
\Else
    \State All users in $U_k$ would join in the auction;
    \State Exit while loop;
\EndIf}
\EndWhile 

\State \textbf{Output:} {\color{black}$p_k \gets \frac{R}{\min\{\sum_{i\in U_k}r_{k,i}, \: v_k \}}$ and  $\textit{Rev}(p_k) \gets R$ with $U_k$ not empty, otherwise, $p_k \gets 0$ and $\textit{Rev}(p_k) \gets 0$.}


\State \textbf{end}
\end{algorithmic}
\end{algorithm}

\begin{theorem}
Mechanism \textbf{iCAT} is dominant strategy incentive compatible.
\label{ICMechanism}
\end{theorem}

\textit{Proof.} If a user rejects an offer, he will be out of the auction and unable to participate in the next round, hence getting a payoff of $0$. Therefore rejecting $p_k$, when $p_k < u_{k,i}$, is a dominated strategy. 

Likewise, accepting $p_k > u_{k,i}$ is a dominated strategy, since prices will rise the next round.
Therefore the dominant strategy for every user $i$ is to report his true value $u_{k,i}$. \qed

The following theorem provides an optimality guarantee for {\color{black}iCAT}. It uses the benchmark $\text{OptRev}^{\geq 2}(U_k^{all}) = \text{max}_{p_k} \: p_k \: \min\{\sum_{i\in U_k^{all}} r_{k,i} \mathds{1}_{\{u_{k,i}\geq p_k\}}, v_k \} $, which has a requirement of at least two users being in the market. This is not a serious constraint in light of the number of users at one base station. {\color{black}Besides, we use $U_k^{all}$ to indicate the initial $U_k$ in \textbf{iCAT}, that is, the total number of users at base station $k$.}

\begin{theorem}
The mechanism \textbf{iCAT} achieves a revenue of $R$ if $\text{OptRev}^{\geq 2}(U_k^{all}) \geq R$, and a revenue of $0$ otherwise.
\end{theorem}

\textit{Proof.} 
According to Theorem \ref{thr:optimalityOPA}, we have
\begin{equation}
\label{eq:optRev}
\text{OptRev}^{\geq 2}(U_k^{all}) = u_{k,x}^* \: \min\left\{\sum_{i\in U_{k,x}^*} r_{k,i}, v_k \right\},
\end{equation}
for some $u_{k,x}^*$ and $U_{k,x}^* = \{i| u_{k,i} \geq u_{k,x}^*\}$. 

{\color{black}If $\text{OptRev}^{\geq 2}(U_k^{all}) > R$, then some $u_{k,x}$ not equal to $u_{k,x}^*$ could be found to obtain a revenue $\textit{Rev}(u_{k,x})$ equal to $R$. On the contrary, if $\text{OptRev}^{\geq 2}(U_k^{all}) < R$, by (\ref{eq:optRev}) we will not be able to find any $u_{k,x}$ satisfying $u_{k,x} \geq \frac{R}{ \min\{\sum_{i\in U_{k}^{all}} r_{k,i}, v_k \}}$. According to line 12 in Algorithm 2, a revenue of 0 is obtained in this case. Besides, for the case with $\text{OptRev}^{\geq 2}(U_k^{all}) = R$, the revenue of $R$ is achieved naturally. \qed}

{\color{black}Intuitively, the target revenue $R$ plays a key role in \textbf{iCAT}. How shall the base station estimate $R$? For truthfulness, we want $R$ to be estimated independently of the bidders we run auction \textbf{iCAT} on. Hence, we further propose a partitioning mechanism \textbf{PUFF} - Partitioning Users For truthFulness, for the base station to estimate $R$ while preserving truthfulness.}

\subsection{Partitioning Users For Truthfulness (PUFF)}

The operations of \textbf{PUFF} are as follows: We partition the set of all users into two sets. Following which, we calculate the optimal revenues $R_1$ and $R_2$ for each set. Next, we use the optimal revenues as 'estimates of $R$' for the opposing set and {\color{black}run \textbf{iCAT} in each set}. Note that when the total supply is less than the total demand, we will run the separate auctions using $\floor{v_k/2}$ and $\ceil{v_k/2}$ number of VMs. 
The complete PUFF is summarized in Algorithm 3.

\begin{algorithm}
\caption{PUFF: Partitioning Users For truthFulness Mechanism}\label{algo:PUFF}
\begin{algorithmic}[1]
\State \textbf{Input:} Initialize $U_k$ and the number of VMs required by user $i$ ($r_{k,i}$).
\State Randomly partition $U_k$ into two sets $S_1$ and $S_2$ \textcolor{black}{of equal size}.
\If{$\sum_{i \in U_k} r_{k,i} > v_k$}
\State Calculate $R_1=$ optimal revenue of $S_1$ given $\floor{v_k/2}$ \hspace*{\algorithmicindent}VMs, and $R_2=$ optimal revenue of $S_2$ given $\ceil{v_k/2}$ \hspace*{\algorithmicindent}VMs; 
\State Run auction \textbf{iCAT}($S_1,\floor{v_k/2}, R_2$) on set $S_1$, and \hspace*{\algorithmicindent}\textbf{iCAT}($S_2,\ceil{v_k/2},R_1$) on set $S_2$. 
\Else 
\State Calculate $R_1=$ optimal revenue of $S_1$ given $v_k$ VMs, \hspace*{\algorithmicindent}and $R_2=$ optimal revenue of $S_2$ given $v_k$ VMs.
\State Run auction \textbf{iCAT}($S_1,v_k, R_2$) on set $S_1$, and \hspace*{\algorithmicindent}\textbf{iCAT}($S_2,v_k,R_1$) on set $S_2$. 
\EndIf
\State \textbf{end}
\end{algorithmic}
\end{algorithm}

In the following theorem, we show that \textbf{PUFF} is truthful.

\begin{theorem}
Mechanism \textbf{PUFF} is dominant strategy truthful.
\end{theorem}
\textit{Proof.} Auction \textbf{iCAT} is truthful when implemented with an $R$ estimated independently of the users it is run on. \qed 

Next, we state a lemma which helps us towards proving lower bounds on the performance of $\textbf{PUFF}$.
\begin{lemma}
The revenue of $\textbf{PUFF}$ is at least $\min(R_1,R_2)$.
\end{lemma}
\textit{Proof.} Either $R_1 > R_2$, $R_2 > R_1$, or $R_1 = R_2$ holds in the \textbf{PUFF}. {\color{black}Therefore, at least one auction out of \textbf{iCAT}($S_1,R_2$) and \textbf{iCAT}($S_2,R_1$) \textcolor{black}{succeeds, i.e. gets a revenue of above 0}, giving a revenue of $\min(R_1,R_2,R_1+R_2)$.} 
\qed

Following which, we prove bounds on the optimality gap of \textbf{PUFF}, proving that its competitive ratio is $4$, {\color{black}in a special case where all users $i$ request one VM, i.e., $r_{k,i}=1$}.

\begin{theorem}
Assume $r_{k,i}=1$ for all users. Let $Rev$ be the expected revenue of \textbf{PUFF}. We will have $\frac{\text{Rev}}{\text{OptRev}^{\geq 2}(U_k^{all})} \geq \frac{1}{4}$. 
\end{theorem}
\textit{Proof.} We know from Theorem \ref{thr:optimalityOPA} that $\text{OptRev}^{\geq 2}(U_k^{all}) = u_{k,x}^{*} \: \min\{\sum_{i\in U_{k,x}^*} r_{k,i}, v_k \}$ for some $u_{k,x}^{*}$ and $U_{k,x}^* = \{i| u_{k,i} \geq u_{k,x}^{*}\}$. {\color{black} Let $D=\sum_{i\in U_k^{all}} r_{k,i}$ and $S=v_k$.} Further, we first analyse the case where $D\geq S$. Given this $u_{k,x}^{*}$, we will have $R_1 \geq u_{k,x} \: \min\{\sum_{i\in U_{k,x}^* \cap S_1} r_{k,i}, \floor{v_k/2} \}$ and $R_2 \geq u_{k,x} \: \min\{\sum_{i\in U_{k,x}^* \cap S_2} r_{k,i}, \ceil{v_k/2} \}$.
Therefore, we deduce that
\begin{equation}
\begin{aligned}
& \frac{Rev}{\text{OptRev}^{\geq 2}(U_k^{all})}
\stackrel{(a)} \geq  \frac{\mathbb{E}[\min(R_1,R_2)]}{ u_{k,x}^{*} \: \min\{\sum_{i\in U_{k,x}^*} r_{k,i}, v_k \}}\\
& \stackrel{(b)} \geq \frac{\mathbb{E}[\min(u_{k,x}^{*} \: \min\{A, \floor{v_k/2} \},u_{k,x}^{*}\: \min\{B, \ceil{v_k/2} \})]}{ u_{k,x}^{*} \: \min\{\sum_{i\in U_{k,x}^*} r_{k,i}, v_k \}}\\
& \stackrel{(c)} \geq \frac{\min(\floor{v_k/2}, \mathbb{E}[ \min\{A, B \}]}{ \min\{\sum_{i\in U_{k,x}^*} r_{k,i}, v_k \}}\\
& \stackrel{(d)} \geq \frac{\min\{ \floor{v_k/2},1/4 \sum_{i\in U_{k,x}^*} r_{k,i}  \}}{\min\{\sum_{i\in U_{k,x}^*} r_{k,i}, v_k \}} \geq \frac{1}{4}.\\
\end{aligned}
\end{equation}
In inequality $(b)$, we have $A=\sum_{i\in U_{k,x}^* \cap S_1} r_{k,i}$ and $B=\sum_{i\in U_{k,x}^* \cap S_2} r_{k,i}$. 
Note that the transition from inequality $(c)$ to $(d)$ is due to the fact that if we flip $k\geq 2 $ coins (corresponding to partitioning the winners into the 2 sets), $\mathbb{E}[\min(H,T)]\geq \frac{1}{4}$  \cite{roughgarden2010algorithmic}, Chapter 13. 

Likewise, for the case where $D \leq S$, following the same logic we have 
\begin{equation}
    \frac{Rev}{\text{OptRev}^{\geq 2}(U_k^{all})} \geq \frac{\min\{ v_k,1/4 \sum_{i\in U_{k,x}^*} r_{k,i}  \}}{\min\{\sum_{i\in U_{k,x}^*} r_{k,i}, v_k \}} \geq \frac{1}{4}. \qed
\end{equation}

It is emphasized that, \textbf{iCAT}, \textbf{PUFF} and their proofs borrow from, but are different from the Profit Extractor and Random Sampling Auction in \cite{goldberg2001competitive,roughgarden2010algorithmic}. Profit Extractor and Random Sampling Auction cater to fully digital goods, with 0 marginal cost of producing the next good, and hence an infinite supply. Unlike this, our network has a limited supply of VMs, resulting in unique novel algorithms and proofs.

\section{Combined Algorithm and Analysis\label{sec:Algorithm_Analysis}}
In this section, we present the combined VM placement and pricing mechanism, describing its implementation. Next, we analyse its performance, \textcolor{black}{termed \textbf{cMAP}}, and prove bounds on the optimality gap caused by potential perturbations \textcolor{black}{on $\Phi^*_{\textbf{v}}$.} 

\subsection{Algorithm Implementation}
The distributed and combined Markov Approx VM Placement and Pricing Algorithm (\textbf{cMAP}) is summarized in Algorithm 4 and works as follows:
\textcolor{black}{Each round, we randomly select a base station.
The base station $k$ considers potential configurations $\textbf{v}'$ in which it has gained one VM, or sent one VM to elsewhere. The network operator obtains the target revenue $\Phi_{\textbf{v}'}$ using \textbf{OPA},\textbf{PUFF}, or via historical data.
The base station then starts exponential clocks for each of these configurations, following the transition rate $q_{\textbf{v}\textbf{v}'} \gets \exp(0.5\beta(\Phi^*_{\textbf{v}'}-\Phi^*_{\textbf{v}}))$.}
When the performance of the target configuration is relatively higher (or lower) than the current, there will be a higher (or lower) rate of switching. The process repeats until convergence to the stationary distribution, the optimal point of \textbf{VP-approx}. This point approximates the optimal point of \textbf{MPP} with an optimality gap of $\frac{1}{\beta}\log|\mathbb{V}|$, according to Proposition \ref{prop:optGap}.
Note that due to its distributed nature, our algorithm is able to handle the dynamic scenarios when new users enter the system, or when users shift from region to region.


\begin{algorithm}
\caption{cMAP: Combined Markov Approx VM Placement and Pricing Algorithm }\label{cMAP} 
\begin{algorithmic}[1]
\State \textbf{Input:} $V$, the total number of VMs across the network, $\{U_k\}$, the set of users across all base stations, and $\{r_{k,i}\}$, {\color{black}the number of VMs required by all users}.
\State Initialise a configuration \textbf{v}.
\State Network operator calculates $\Phi^*_{\textbf{v}} \leftarrow$ \textbf{OPA}(\textbf{v},$\{U_k\}$, $\{r_{k,i}\}$) or \textbf{PUFF}(\textbf{v},$\{U_k\}$, $\{r_{k,i}\}$);
\While{True}
\State Randomly select a base station $k$. 
\State \textcolor{black}{Consider configurations $\textbf{v}'$ with $v_k \pm 1$ VMs at $k$.}
\For{\textcolor{black}{all configurations $\textbf{v}'$}}
\State Network operator obtains the target revenue $\Phi^*_{\textbf{v}'} \leftarrow$ \hspace*{\algorithmicindent}\hspace*{\algorithmicindent}\textbf{OPA}(\textbf{v}$'$,$\{U_k\}$, $\{r_{k,i}\}$) or \textbf{PUFF}(\textbf{v}$'$,$\{U_k\}$, $\{r_{k,i}\}$);
\State Set clocks with transition rate $q_{\textbf{v}\textbf{v}'}\gets$ \hspace*{\algorithmicindent}\hspace*{\algorithmicindent}$\exp(0.5\beta(\Phi^*_{\textbf{v}'}-\Phi^*_{\textbf{v}}))$;
\EndFor
\State The CTMC transits to the next state according to $q_{\textbf{v}\textbf{v}'}$;
\EndWhile
\end{algorithmic}
\end{algorithm}

\subsection{Algorithm Analysis}
Our combined mechanism \textbf{cMAP} attains an optimality gap of $\frac{1}{\beta}\log|\mathbb{V}|$ for the original problem \textbf{MPP}.
In practice, the system may obtain an inaccurate value of $\Phi_{\textbf{v}}^*$, the optimal revenue under configuration $\textbf{v}$. This may occur when we implement the incentive compatible auction mechanism \textbf{PUFF} and estimate $R$.

In light of this we analyse the impact of the perturbations, by bounding the optimality gap caused by the perturbations, on problem $\textbf{VP-approx}$. To this end, we construct a new CTMC which takes into account the perturbations, and characterize its stationary distribution, in the following.

For each state $\textbf{v}$ with optimal revenue $\Phi_{\textbf{v}}^*$, we let $\overline{\Phi}_{\textbf{v}}$ be its corresponding perturbed inaccurate revenue. The perturbation error $\epsilon_{\textbf{v}} = \overline{\Phi}_{\textbf{v}}-  \Phi_{\textbf{v}}^*$ lies in the range $[-\psi_{\textbf{v}}, \psi_{\textbf{v}}]$. 
For each state $\textbf{v}$, we quantize the error into $2a_{\textbf{v}}+1$ potential values $[-\psi_{\textbf{v}},... , -\psi_{\textbf{v}}/a_{\textbf{v}}, 0, ..., \psi_{\textbf{v}}/a_{\textbf{v}}, ..., \psi_{\textbf{v}}]$, where the error $\epsilon_{\textbf{v}} = \frac{n}{a_{\textbf{v}}}\psi_{\textbf{v}}$ with probability $\rho_{\textbf{v}_n}, n=0,\pm 1, .. \pm a_{\textbf{v}}$, and $\sum_{n=-a_{\textbf{v}}}^{a_{\textbf{v}}}\rho_{\textbf{v}_n}=1$. This means that we have constructed a new CTMC in which each state ${\textbf{v}}$ of the original CTMC is now expanded into $2 a_{\textbf{v}} + 1$ states. The transition rate follows the following equation:
\begin{equation}
    q_{\textbf{v}_n\textbf{v}_{n'}^{'}}= \exp(0.5 \beta (\Phi_{\textbf{v}'_{n'}}^* -\Phi_{\textbf{v}_n}^*)) \rho_{\textbf{v}'_{n'}}.
\end{equation}
Based on the detailed balanced equation $\pi_{\textbf{v}_n} q_{\textbf{v}_n \textbf{v}'_{n'}}= \pi_{\textbf{v}'_{n'}}  q_{\textbf{v}'_{n'}\textbf{v}_n}$, we have 
\begin{equation}
    \pi_{\textbf{v}_n}\! \exp(\frac{1}{2}\beta (\Phi_{\textbf{v}'_{n'}}^* -\Phi_{\textbf{v}_n}^*)) \rho_{\textbf{v}'_{n'}} \!=\!\pi_{\textbf{v}'_{n'}}\! \exp(\frac{1}{2}\beta (\Phi_{\textbf{v}_{n}}^* -\Phi_{\textbf{v}'_{n'}}^*)) \rho_{\textbf{v}_{n}},
\end{equation}
which results in
\begin{equation}
    \pi_{\textbf{v}_n} \exp(\beta \Phi_{\textbf{v}'_{n'}}^*) \rho_{\textbf{v}'_{n'}} =\pi_{\textbf{v}'_{n'}} \exp(\beta \Phi_{\textbf{v}_{n}}^*) \rho_{\textbf{v}_{n}}.
\end{equation}
Because $\sum_{\textbf{v}' \in \mathbb{V}} \sum_{n'=-a_{\textbf{v}'}}^{a_{\textbf{v}'}} \pi_{\textbf{v}'_{n'}}=1$, we obtain 
\begin{equation}
    \pi_{\textbf{v}_n} =\frac{ \exp(\beta \Phi_{\textbf{v}_{n}}^*) \rho_{\textbf{v}_{n}}}{\sum_{\textbf{v}' \in \mathbb{V}} \sum_{n'=-a_{\textbf{v}'}}^{a_{\textbf{v}'}} \exp(\beta \Phi_{\textbf{v}'_{n'}}^*) \rho_{\textbf{v}'_{n'}} }.
\end{equation}
Letting $\sigma_{\textbf{v}'}=\sum_{n'=-a_{\textbf{v}'}}^{a_{\textbf{v}'}} \rho_{\textbf{v}'_{n'}}\exp(\beta \frac{n'}{a_{\textbf{v}'}}\psi_{\textbf{v}'})$, the distribution of the new perturbed CTMC will be 
\begin{equation}
\label{eq:PerturbedDist}
    \overline{\pi}_{\textbf{v}} = \sum_{n=-a_{\textbf{v}}}^{a_{\textbf{v}}}\pi_{\textbf{v}_n} = \frac{\sigma_{\textbf{v}} \exp(\beta \Phi^*_{\textbf{v}})}{\sum_{\textbf{v}' \in \mathbb{V}} \sigma_{\textbf{v}'} \exp(\beta \Phi^*_{\textbf{v}'})}.
\end{equation}

We use the Total Variation Distance \cite{diaconis1991geometric, zhang2013optimal} as a metric to quantify the optimality gap between the stationary distribution of the perturbed CTMC $\overline{\pi}_{\textbf{v}}$ and $\pi_{\textbf{v}}^*$, the optimal solution of $\textbf{VP-approx}$, {\color{black}as follows:}
\begin{equation}
    d_{TV}(\pi_{\textbf{v}}^*, \overline{\pi}_{\textbf{v}})=\frac{1}{2}\sum_{\textbf{v}\in \mathbb{V}} |\pi_{\textbf{v}}^*-\overline{\pi}_{\textbf{v}}|.
\end{equation}
With the stationary distribution of the perturbed CTMC $\overline{\pi}_{\textbf{v}}$, we use a result in \cite{zhang2013optimal}, which proved that the total variation distance is bounded as follows
\begin{equation}
    d_{TV}(\pi_{\textbf{v}}^*, \overline{\pi}_{\textbf{v}}) \leq 1-\exp(-2\beta \psi_{\text{max}}),
\end{equation}
where $\psi_{\text{max}}=\text{max}_{\textbf{v}} \psi_{\textbf{v}}$, the largest perturbation error among states $\textbf{v}$. The revenue gap is hence bounded as follows:
\begin{equation}
    |\pi_{\textbf{v}}^* \Phi^*_{\textbf{v}}- \overline{\pi}_{\textbf{v}} \overline{\Phi}_{\textbf{v}}| \leq 2 \Phi_{\text{max}} (1-\exp(-2\beta \psi_{\text{max}})),
\end{equation}
where $\Phi_{\text{max}}=\text{max}_{\textbf{v}} \Phi_{\textbf{v}}$.

The upper bound on both the Total Variation Distance between the two distributions $d_{TV}(\pi_{\textbf{v}}^*, \overline{\pi}_{\textbf{v}})$ and the optimality gap $|\pi_{\textbf{v}}^* \Phi^*_{\textbf{v}}- \overline{\pi}_{\textbf{v}} \overline{\Phi}_{\textbf{v}}|$ is independent with respect to $\rho_{\textbf{v}_n}$, the distribution of perturbed revenues, and is independent with respect to $|\mathbb{V}|$, the total number of configurations. This indicates that the optimality gap does not increase with the network size and number of \textcolor{black}{configurations $|\mathbb{V}|$. Besides this, using Markov Approximation enables us to perform a distributed implementation on this large combinatorial problem.} 


\section{Simulation Results\label{sec:Simulations}}
In this section, we evaluate the performance of {\color{black}our combined mechanisms \textbf{cMAP}: \textbf{MAP} (which solves the VM placement problem) along with either \textbf{OPA} or \textbf{PUFF} (which solve the normalized pricing problem)}, and provide some insights. 

\subsection{Convergence, and insights on pricing}
Firstly, we consider a network in which there are 5 BSs, and 10 VMs being distributed amongst these 5 BSs. The 5 base stations have $(2,0,2,4,0)$ users respectively. We set $r_{k,i}$, the number of VMs required by user $i$ at BS $k$, to be between 1 to 3 VMs. $u_{k,i}$, the willingness to pay of user $i$ at BS $k$, follows uniform distribution $U[0,1]$.
\begin{figure}[ht]
\centering
\includegraphics[angle=0,scale=0.5]{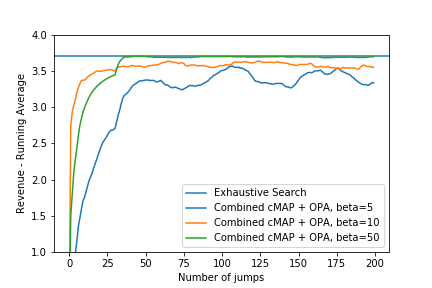}
\caption{Convergence of the \textbf{cMAP}. 
}
\label{fig:Convergence}
\end{figure}
\setlength{\textfloatsep} {3pt}

Under this setup, we run \textbf{cMAP} (the combined Markov Approximation VM Placement Algorithm) along with auction \textbf{OPA}. 
for different values of $\beta$. We plot the running average over a window size of $30$ jumps, in comparison with the optimal value, as seen in Fig \ref{fig:Convergence}. The optimal value is obtained by exhaustive search, evaluating the solution to $\textbf{MPP}$ over all combinations of $\textbf{v}$.
As seen, for $\beta=50$, we are able to achieve optimality. For $\beta=10$, the converged stationary distribution over configurations of $\textbf{v}$ is near optimal. 
Under $\beta=10$, the top 5 most common states are $\textbf{v}=(2,0,2,5,1)$, $(2,0,3,5,0)$, $(2,1,2,5,0)$, $(3,0,2,5,0)$, $(2,0,2,6,0)$, which are best able to meet the total demand of $(2,0,2,4,0)$.
Notice that as $\beta$ increases, performance improves: the running average is closer to the optimal point, and fluctuations decrease.
The fluctuations occur because under our Markov Approximation-inspired algorithm, we converge not to a specific state of the CTMC, but to a stationary distribution over the states of the CTMC. Recall that the converged stationary distribution has an optimality gap of $\frac{1}{\beta} \log|\mathbb{V}|$ from the optimal point of the original problem \textbf{VP}. This shows that as $\beta \rightarrow \infty$, the performance converges to the optimal value of \textbf{VP}.
A potential tradeoff in having a higher $\beta$ is: if $\Phi_{\textbf{v}}^*>\Phi_{\textbf{v}'}^*$, according to (\ref{eq:transitionEq}) there will be a lower rate of switching, and a higher probability of staying in the current state. As $\beta$ increases, the network is more likely to stay in the current state. This may lead to a longer time spent in local minimums, due to the lack of exploration, and hence a longer convergence time.

Next, with the current setup we compare the performance of our proposed mechanisms \textbf{MAP} + \textbf{OPA} 
and \textbf{MAP} + \textbf{PUFF} 
to the following baselines:

1) Cooperative BS + Uniform Pricing: Under this scenario, the base stations are cooperative. They share the VMs with each other, where the VMs are transferred within the network via our proposed \textbf{MAP}. Unlike our proposed combined solution, here we use uniform pricing: a common price is set throughout the network, regardless of the demand pattern. A benefit of uniform pricing is that it is faster to implement.

2) Non-cooperative BS + Auction: Under this scenario, the base stations are no longer cooperative - they do not share the VMs with each other. We obtain the average result under the non-cooperative scenario, by averaging over all the possible combinations of $\textbf{v}$. For each configuration $\textbf{v}$, we use the optimal auction \textbf{OPA} to obtain $\Phi_{\textbf{v}}^*$.

3) Non-cooperative BS + Uniform Pricing: {\color{black}Under this scenario, the base stations not only do not share the VMs with each other, but also do not consider the demand pattern, using a common price throughput the network.}
\begin{figure}[ht]
\centering
\includegraphics[angle=0,scale=0.5]{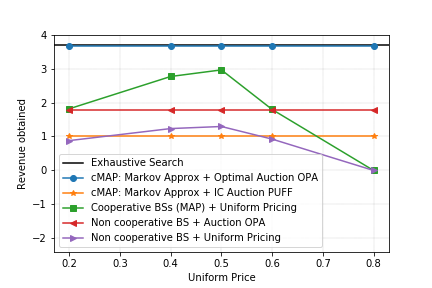}
\caption{The effect of different uniform prices on revenue.
}
\label{fig:unifPriceComp}
\end{figure}
\setlength{\textfloatsep} {3pt}

We plot the revenue obtained under the various methods, and show how the performance varies when different prices are set as the uniform price in Fig. \ref{fig:unifPriceComp}.
As seen, our proposed algorithms \textbf{cMAP} outperforms the baselines, especially when \textbf{OPA} is used as the pricing mechanism. While \textbf{MAP} in combination with \textbf{PUFF} is not near-optimal, we have proved that \textbf{PUFF} has a competitive ratio of $4$.
The baselines involving uniform price perform best when the price is "neutral" - neither too low nor too high. If the price is too high, the users (likely having a lower willingness to pay) would not choose to use the VMs. If the price is too low, the revenue the network operator obtains will be low.
Fig. \ref{fig:unifPriceComp} also shows that resource sharing among base stations increases the revenue.

\subsection{A larger setup, with insights on willingness to pay and the demand-supply ratio}
Next, we enlarge our setup and compare the performance of our proposed mechanisms with the different baselines.
In this setup, there are 20 VMs shared amongst the 5 base stations. The number of users at each base station are randomized, along with $r_{k,i}$, the number of VM units each user requests. 
We let the users' willingness to pay $u_{k,i}$ follow a uniform distribution $U[a,b]$.
\begin{figure}[ht]
\centering
\includegraphics[angle=0,scale=0.5]{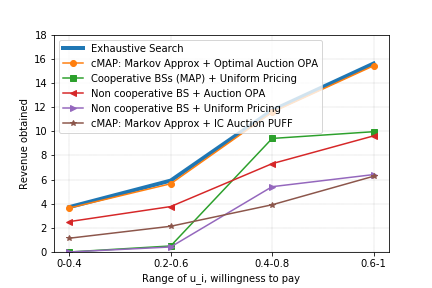}
\caption{The impact of willingness to pay on revenue. 
}
\label{fig:varyUi}
\end{figure}
\begin{figure}[ht]
\centering
\includegraphics[angle=0,scale=0.5]{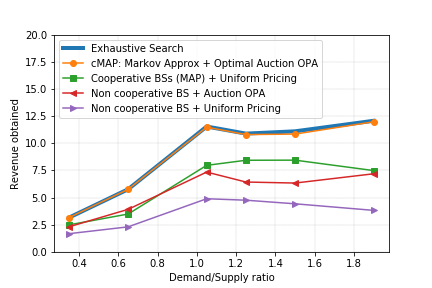}
\caption{The impact of the Demand/Supply ratio on revenue. 
}
\label{fig:varySDratio}
\end{figure}

In Fig. \ref{fig:varyUi}, we show the impact of users' willingness to pay on the revenue. The range of $u_{k,i}$ is adjusted, from the uniform distribution $U[0,0.4]$ (low willingness to pay), to $U[0.2,0.6]$, $U[0.4,0.8]$ and $U[0.6,1]$ (high willingess to pay). Our propsed solution \textbf{MAP} + \textbf{OPA} (with $\beta=10$) outperforms the baselines, obtaining a near-optimal revenue.
Our results show that on average, having base station cooperation increases the revenue by up to $53 \%$ percent.
As seen in Fig. \ref{fig:varyUi}, when the users have a higher willingness to pay, the revenue increases.
Notice that uniform pricing ($p=0.5$) does not perform well, when the users have low willingness to pay.

Fig. \ref{fig:varySDratio} illustrates the impact of revenue when the $\frac{\text{Demand}}{\text{Supply}}$ ratio is varied. Supply is fixed at $20$ VMs, while demand is increased, from $D=7$ (low demand), to $D=21$ (near equal demand and supply) and high demand $D=38$.
Our solution \textbf{cMAP} outperforms the baselines, especially when demand increases, as the supply of VMs is shifted around the network to meet demand more effectively, and {\color{black}an optimal auction} is used to extract the highest revenue possible.
Our results show that on average, having base station cooperation increases the revenue by up to $57 \%$.
As seen in Fig. \ref{fig:varySDratio}, as the $\frac{\text{Demand}}{\text{Supply}}$ ratio increases, revenue increases because more units of demand are being met. Once the $\frac{\text{Demand}}{\text{Supply}}$ ratio hits 1, revenue no longer increases much due to the lack of global supply in the system.

\section{Conclusions\label{sec:Conclusions}}
{\color{black}In this paper, we have addressed the load-unbalanced problem in MEC systems, by jointly optimizing the VM placement and pricing across base stations. Specifically, we have formulated a revenue maximization problem from the network operator's perspective, which was decomposed to a VM placement master problem and a normalized pricing slave problem. The objective function of the master problem is the optimal value of the slave problem. Then, we solved the master problem by designing a CTMC and solved the slave problem by proposing auctions considering users' truthful and untruthful behaviors, respectively. By combining the algorithms proposed for the master and slave problem, \textbf{cMAP} is implemented for VM placement and pricing decision making across base stations. Through theoretical analysis, we give the optimal gap of \textbf{cMAP}, which is nested with \textbf{OPA} (auction mechanism with users' truthful behaviors) and \textbf{PUFF} (auction machanism with users' untruthful behaviors), respectively. Finally, we demonstrated the convergence and efficiency of \textbf{cMAP}.}
For future work we will analyse the impact of factors like having a heterogeneous cost of VM migration between base stations.




\bibliographystyle{IEEEtran}

\bibliography{referrences} 

\ifCLASSOPTIONcaptionsoff
  \newpage
\fi

\end{document}